\documentclass[preprint]{aastex}
\usepackage{lscape,morefloats}

\begin{document}

\title{The Neptune-Sized Circumbinary Planet Kepler-38b$^{\dagger}$}

\author{Jerome A. Orosz\altaffilmark{1}, 
William F. Welsh\altaffilmark{1},
Joshua A. Carter\altaffilmark{2,3},
Erik Brugamyer\altaffilmark{4},
Lars A. Buchhave\altaffilmark{5,6},
William D. Cochran\altaffilmark{4},
Michael Endl\altaffilmark{4}, 
Eric B. Ford\altaffilmark{7},
Phillip MacQueen\altaffilmark{4}, 
Donald R. Short\altaffilmark{1}, 
Guillermo Torres\altaffilmark{2},
Gur Windmiller\altaffilmark{1},
Eric Agol\altaffilmark{8},
Thomas Barclay\altaffilmark{9,10}, 
Douglas A. Caldwell\altaffilmark{9,11}, 
Bruce D. Clarke\altaffilmark{9,11},  
Laurance R. Doyle\altaffilmark{11},
Daniel C. Fabrycky\altaffilmark{3,12}
John C. Geary\altaffilmark{2}, 
Nader Haghighipour\altaffilmark{13},
Matthew J. Holman\altaffilmark{2}, 
Khadeejah A. Ibrahim\altaffilmark{9,14},
Jon M. Jenkins\altaffilmark{9,11}, 
Karen Kinemuchi\altaffilmark{9,10}, 
Jie Li\altaffilmark{9,11},
Jack J. Lissauer\altaffilmark{9},
Andrej Pr\v{s}a\altaffilmark{15},
Darin Ragozzine\altaffilmark{2,7},
Avi Shporer\altaffilmark{16,17,18},
Martin Still\altaffilmark{9,10},
Richard A. Wade\altaffilmark{19}
}
\altaffiltext{1}{Department of Astronomy, San Diego State University,
5500 Campanile Drive, San Diego, CA 92182}
\altaffiltext{2}{Harvard-Smithsonian Center for Astrophysics, 60 Garden Street,
  Cambridge, MA 02138}
\altaffiltext{3}{Hubble Fellow}
\altaffiltext{4}{McDonald Observatory, The University of 
   Texas at Austin, Austin, TX 78712-0259}
\altaffiltext{5}{Niels Bohr Institute, University of Copenhagen, 
   Juliane Maries vej 30, 2100 Copenhagen,  Denmark}
\altaffiltext{6}{Centre for Star \& Planet Formation, Natural History 
      Museum of Denmark, University of Copenhagen,
     \O ster Voldgade 5-7,  1350 Copenhagen, Denmark}
\altaffiltext{7}{Astronomy Department, University of Florida, 211
  Bryant Space Sciences Center, Gainesville, FL 32111}
\altaffiltext{8}{Department of Astronomy, BOX 351580, University
  of Washington, Seattle, WA 98195}
\altaffiltext{9}{NASA Ames Research Center, M/S 244-40, Moffett Field, CA
          94035}
\altaffiltext{10}{Bay Area Environmental Research Institute, Inc.,
       560 Third Street West, Sonoma, CA 95476}
\altaffiltext{11}{SETI Institute, 189 Bernardo Avenue, Mountain View, CA 94043}
\altaffiltext{12}{Department of Astronomy \& Astrophysics, 
  University of California,
  Santa Cruz, Santa Cruz, CA 95064}
\altaffiltext{13}{Institute for Astronomy and NASA Astrobiology Institute
  University of Hawaii-Manoa, 2680 Woodlawn Dr., Honolulu, HI 96822}
\altaffiltext{14}{Orbital Sciences Corporation/NASA Ames Research Center, 
  Moffett Field, CA 94035}
\altaffiltext{15}{Department of Astronomy and Astrophysics, 
  Villanova University, 800 E Lancaster Avenue, Villanova, PA 19085}
\altaffiltext{16}{Las Cumbres Observatory Global Telescope 
   Network, 6740 Cortona 
   Drive, Suite 102, Santa Barbara, CA 93117, USA}
\altaffiltext{17}{Department of Physics, Broida Hall, 
   University of California, Santa Barbara, CA 93106, USA}
\altaffiltext{18}{Division of Geological and Planetary 
Sciences, California Institute
of Technology, Pasadena, CA 91125}
\altaffiltext{19}{Department of Astronomy \& Astrophysics, 
       The Pennsylvania State University, 525 Davey Lab, 
       University Park, PA 16802}
\altaffiltext{$\dagger$}{Based on 
 observations obtained with the Hobby-Eberly Telescope, 
 which is a joint project of the 
 University of Texas at Austin, 
 the Pennsylvania State University, 
 Stanford University, Ludwig-Maximilians-Universit\"at M\"unchen, and 
 Georg-August-Universit\"at G\"ottingen.}

\begin{abstract}
We discuss the discovery and characterization of the circumbinary
planet Kepler-38b.  The 
stellar binary is single-lined, with a period of 18.8 days, and
consists of a moderately  evolved main-sequence 
star ($M_A=0.949\pm 0.059\,M_{\odot}$ and $R_A=1.757\pm 0.034\,R_{\odot}$)
paired with a low-mass star 
($M_B=0.249\pm 0.010\,M_{\odot}$ and $R_B=0.2724\pm 0.0053\,R_{\odot}$)
in a mildly eccentric ($e=0.103$) orbit.
A total of eight transits due to a circumbinary planet
crossing the primary star
were identified
in the {\em Kepler} light curve (using {\em Kepler}
Quarters 1 through 11),
from which a planetary period of $105.595\pm 0.053$ days
can be established.  
A photometric dynamical model fit to the 
radial velocity curve and
{\em Kepler} light curve 
yields a planetary radius of $4.35\pm 0.11\,R_{\oplus}$, or
$1.12\pm 0.03\,R_{\rm Nep}$.  Since the planet 
is not sufficiently
massive to observably alter the orbit of the binary 
from Keplerian motion,
we can only place
an upper limit on the mass of the planet of $122\,M_{\oplus}$
($7.11\,M_{\rm Nep}$ or 
$0.384\,M_{\rm Jup}$) at 95\% confidence.  
This upper limit should decrease as more {\em Kepler}
data become available.  
\end{abstract}

\section{Introduction}\label{Introduction}

While the {\it Kepler} Mission (Borucki et al.\ 2010) 
is sometimes considered synonymous with
``the search for Earth-like planets'', 
its goals are considerably broader, and include estimating the frequency 
and orbital distribution of planets in multiple-stellar systems. 
To achieve its goals, {\it Kepler} relies on 
its exquisite photometric precision, 
its ability to simultaneously observe roughly 160,000 stars, and 
its long-duration and near-continuous time series measurements
(Koch et al.\ 2010).
This triad of unique capabilities makes {\it Kepler} ideally suited 
for exoplanet discovery and characterization, including planets in
binary star systems [see Haghighipour (2010) for an in-depth 
discussions of planets in 
binary star systems].

If the binary star's orbital plane is favorably oriented,
the stars will eclipse and thus reveal their binary nature.
{\it Kepler} has discovered over 2000 eclipsing binaries
(Pr\v sa et al.\ 
2011; Slawson
et al.\ 2011), with periods ranging from 0.075 to over 275 
days, and these systems are being searched for the presence of planets.
The eclipses tell us that we are viewing the binary system close to 
its
orbital plane, and thus perhaps at a favorable orientation for finding 
transiting planets if the planets lie in the same orbital plane.
However, detecting such planets is much more difficult than finding 
planets orbiting a single star. A dilution factor is present, but the main 
challenges arise from the fact that the transits are neither periodic
nor equal in duration (e.g.\ see Doyle et al.\ 2011; Welsh et al.\ 2012). 
In addition,
the deep stellar
eclipses can easily mask a small transit signal.
Partially compensating for these disadvantages, the timing of the eclipses
of the binary component
stars provides a very sensitive indicator of the presence
of a third body in the system (e.g.\ Orosz et al.\ 2012).
The eclipse timing variations (ETVs) as seen in an O-C diagram
(Observed-minus-Computed) can reveal deviations from periodicity
that are attributed to a 
gravitational perturbation caused by a planet.
Note that for short orbital-period binaries, 
the ETVs are generally dominated by dynamical effects, not 
light-travel time delays.

The first transiting circumbinary planet discovered
was Kepler-16b
(Doyle et al.\ 2011). The transits left no room for ambiguity as to 
the
planetary nature of the third object. 
The planet is in a P-type orbit (Dvorak 1984, 1986),
meaning the planet is circumbinary (an outer orbit around both stars).
Soon after, two more transiting circumbinary planets were discovered,
Kepler-34b and Kepler-35b (Welsh et al.\ 2012), establishing that 
such planets are not rare. While there is considerable diversity among
the three systems (in mass ratios, eccentricities, orbital periods),
two features are in common:
(i) all three planets have a radius similar to
Saturn's, which is interesting in that 
Jupiter-radius planets should be easier to detect; and 
(ii) the orbital periods of the planets are only slightly longer than 
the minimum needed to guarantee dynamical stability according to the
criteria given in
Holman \& Wiegert (1999). 
Whether this is a consequence of planet formation and migration, or simply 
a selection effect, is unknown.

In this paper we announce the discovery of a fourth transiting circumbinary
planet, Kepler-38b. As with the other cases, the detection was made by
visual inspection of a subset of the eclipsing binary star light curves,
namely, those with orbital period greater than $\sim$1 day. The 
observations are presented in \S2, and the photometric-dynamical model
fit in \S3.  We conclude with a discussion in
\S4.

\section{Observations}

\subsection{{\em Kepler} Light Curves}

The details of the {\em Kepler}\, mission have been presented
in \citet{Borucki_2010}, \citet{Koch_2010},
\citet{Batalha_2010},
\citet{Caldwell_2010}, and
\citet{Gilliland_2010} and references therein.
The {\em Kepler} reduction pipeline 
\citep{Jenkins_2010a,Jenkins_2010b}
provides two types of
photometry, the basic simple
aperture photometry (SAP), and the 
``pre-search data conditioned'' (PDC)
data in which
many of the instrumental trends are removed
\citep{Smith_2012,Stumpe_2012}.  
Using the PDC 
light curves through Q11, 
we conducted a visual search for small transit events
in the long-period eclipsing binaries from the catalog of
Slawson et al.\ (2011).  Several candidate systems were found, 
including Kepler-38 (KIC 6762829, KOI-1740,
2MASS J19071928+4216451).  The nominal stellar parameters listed
in the Kepler Input Catalog (KIC, Brown et al.\ 2011) are
a temperature of $T_{\rm eff}=5640$~K, a surface
gravity of $\log g=4.47$, and apparent magnitudes of $r=13.88$
and $Kp=13.94$.
The binary period is 18.8 days, and the eclipses are relatively shallow.
The depth of the primary eclipse is $\approx 3\%$, and the
depth of the secondary eclipse (which is total) is $\approx 0.1\%$.

\begin{figure}
\includegraphics[scale=0.65,angle=-90]{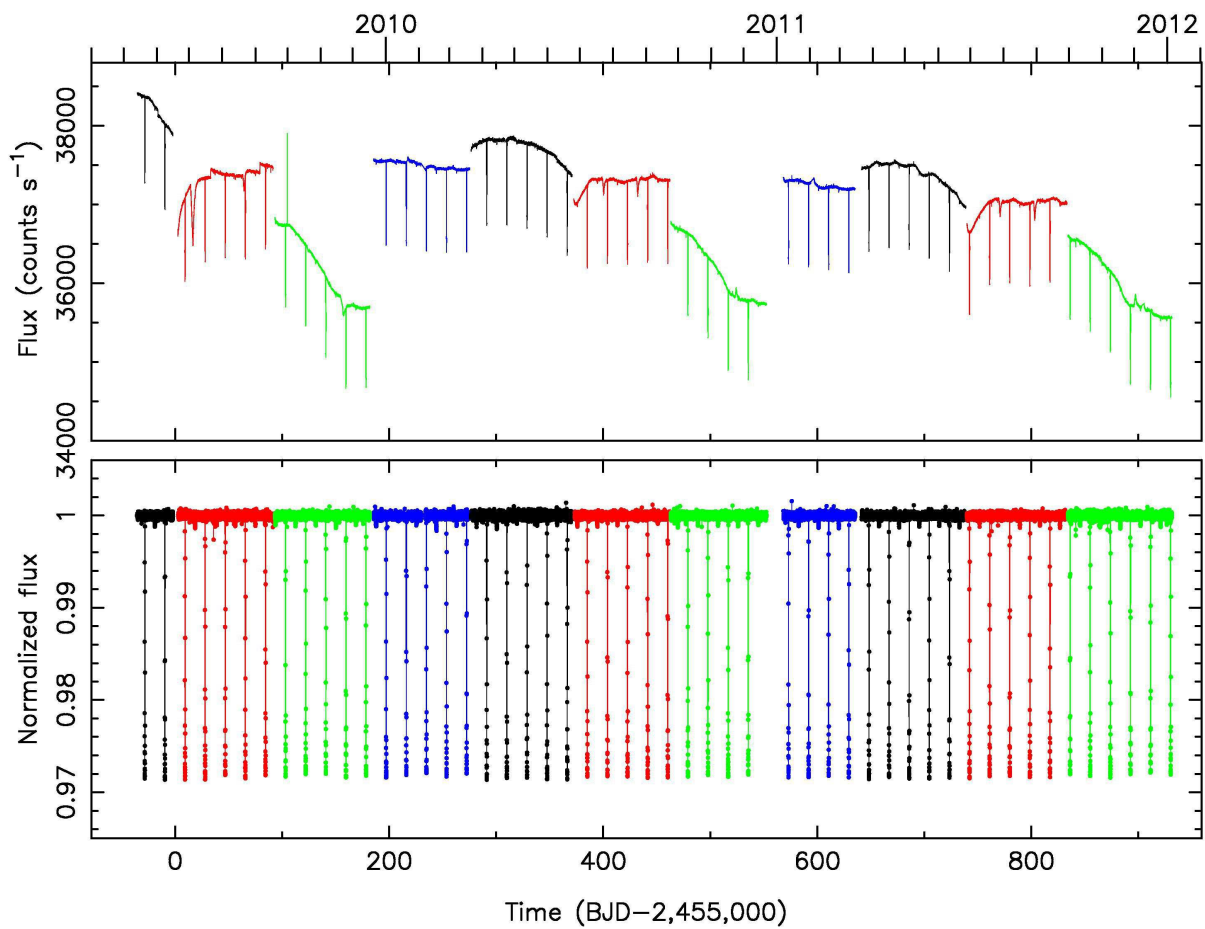}
\caption{Top: the SAP light curves of Kepler-38.
The colors denote the season and hence the spacecraft
orientation with black for Q1, Q5, and Q9,
red for Q2, Q6, and Q10, green for Q3, Q7,
and Q11, and blue for Q4 and Q8.
Bottom: The normalized light curve with the instrumental
trends removed.  One primary eclipse was missed in the
relatively long interval between the end of
Q7 and the start of Q8.
\label{plotraw}}
\end{figure}

While the PDC detrended light curves are convenient for visual
searches, the detrending is not always complete.
In addition, the PDC process attempts to correct for light from 
contaminating sources in the aperture, and correct for light lost
from the target that falls outside the aperture.
We have found that too much correction is sometimes applied, so 
we therefore use the SAP (long cadence only) light curves
for the analysis that is described below.  The SAP 
light curve
was detrended in a manner similar to what Bass et al.
(2012) used.  Briefly, each Quarter was treated as an independent
data set.
The light curve was further divided up into separate
segments, using discontinuities and data gaps as end points.
Splines were fit to each segment, where 
the eclipses and transits were masked out using an iterative
sigma-clipping routine.   Once satisfactory fits were found,
the data were then normalized by the splines and the segments
were combined.
Figure
\ref{plotraw} shows the SAP and normalized 
light curves.  The data span
966.8 days (2.65 years) 
from 2009 May to 2012 January.  The {\em Kepler}
spacecraft was collecting data 92.3\% of the time,
and in the case of Kepler-38, 93.8\% of the observations collected
were flagged as normal (FITS keyword SAP\_QUALITY=0).

\begin{figure}
\centerline{\includegraphics[scale=0.85,angle=0]{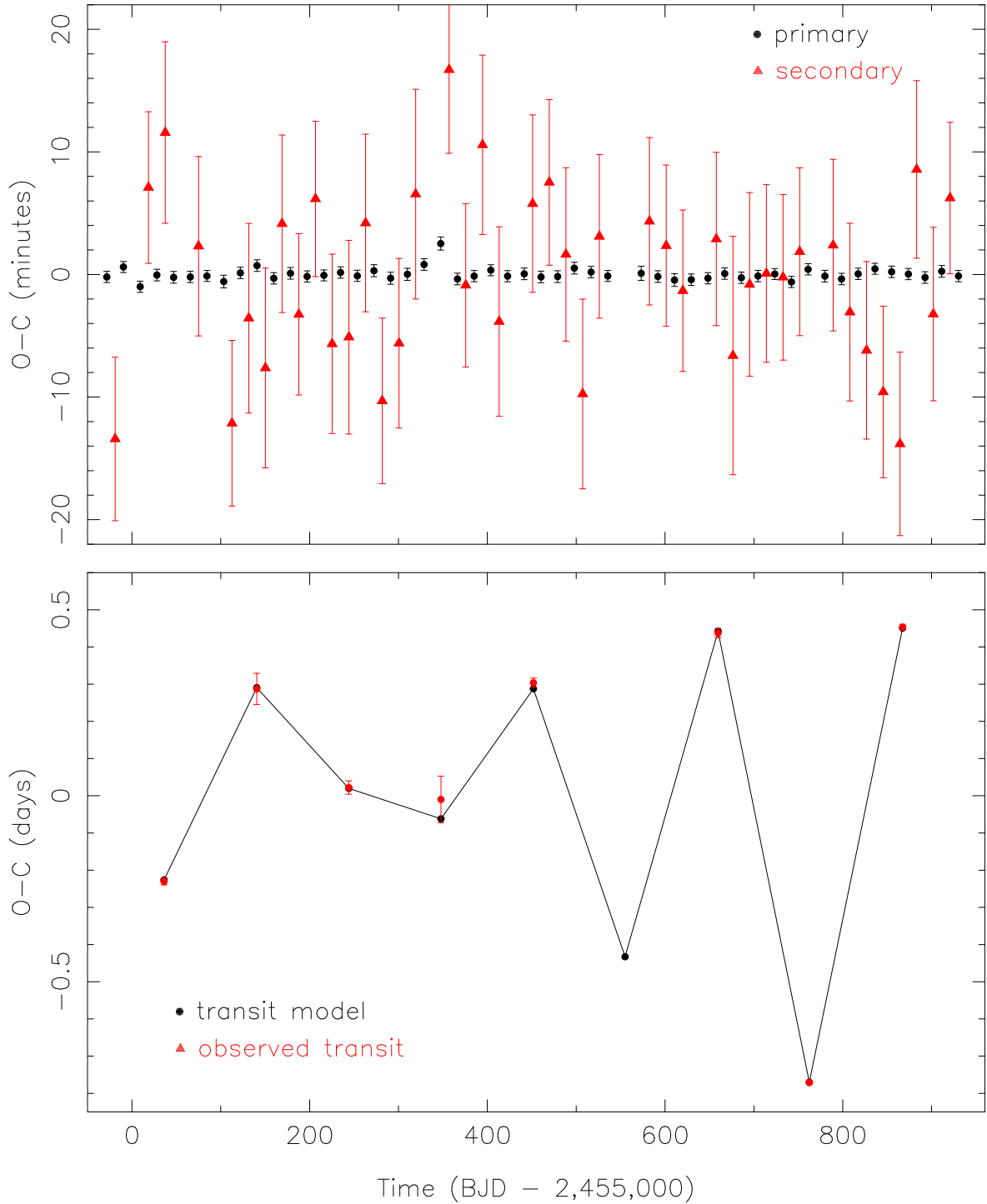}}
\caption{Top: O-C curves for the primary eclipses
(filled circles) and secondary eclipses (filled triangles). 
The units on the vertical scale are minutes.  No significant
trends are evident.
Bottom:  O-C curve for the transits.  Here, the units
on the vertical scale
are days.  The large deviations from a linear
ephemeris rule out a background eclipsing binary.
\label{plotOC}}
\end{figure}

\begin{figure}
\includegraphics[scale=0.7,angle=-90]{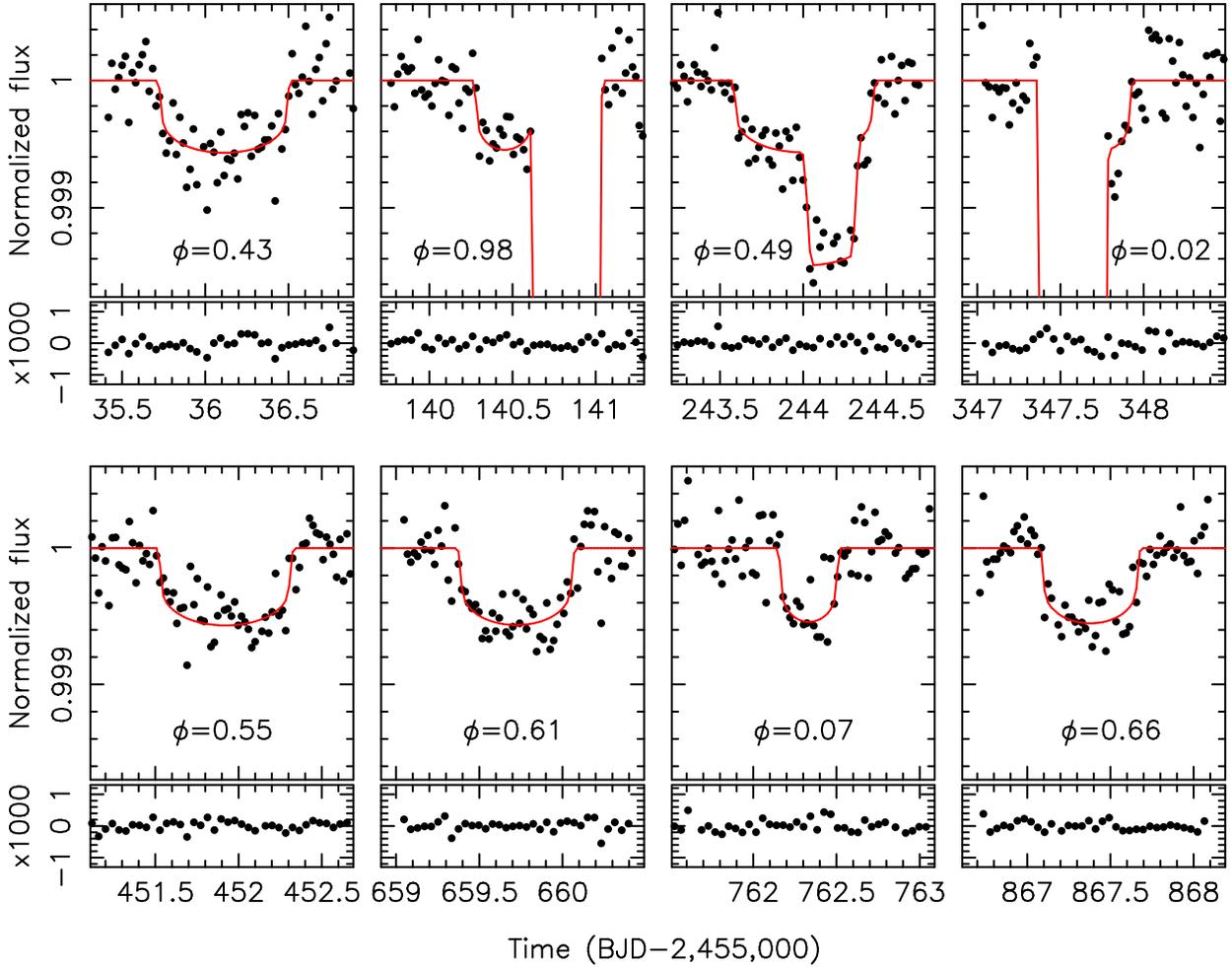}
\caption{The unbinned {\em Kepler}
light curves with the eight transit events
and the best-fitting model are shown.
The orbital phase of each
event is indicated.  Note
the correlation between the width of the transit
event and the orbital phase.  Transits near primary
eclipse ($\phi=0$) are narrow, whereas transits
near the secondary eclipse ($\phi\approx0.5$) are
wide.
\label{koi1740_transits}}
\end{figure}

Initially, six transit events were noticed in the light curve.  
These are all due to a small body transiting the primary star.
The times of mid-transit, the widths, and the depths of the transits
were measured by fitting a simple ``U-function'' (a symmetric low-order 
polynomial)
and the
results are given in Table 1.  The mean period is 103.8 days.  Note,
however, that
the transit times are poorly 
described by a simple linear ephemeris
(see Figure \ref{plotOC}). Furthermore, the widths of the transits
depend on the binary phase, where transits that are 
closer to 
a primary
eclipse (phase $\phi=0$) are narrower than the average
and transits that are 
closer to secondary eclipses
($\phi\approx 0.5$)
are wider than the average.  The large variations in the transit
times and widths are a clear signature of a circumbinary object,
rather than a background eclipsing binary.
After a more detailed analysis was performed, two additional transits
that were partially blended with primary eclipses were found, bringing
the total number of transits to eight.  Another transit was missed 
during the data download between  Q7 and Q8.
Figure \ref{koi1740_transits}  shows the transits and the best-fitting
model, which is described in Section 3.1.  No 
convincing signatures of the transits of the planet across
the secondary were found
in the light curve, and this point is discussed further in
Section 3.1.

We also applied a novel 
transit search algorithm, 
the ``Quasi-periodic 
Automated Transit 
Search'' 
(QATS, Carter \& Agol, 2012), which allows for slight variations in 
the inter-transit spacing. We searched through a range of periods from 
50 to 300 days, and for each trial period, we computed the transit 
duration expected at each time in the light curve for an edge-on 
circular-orbiting planet.  We detrended the light curve and
convolved it with a boxcar with the given
transit duration, and then shifted each time by the duration 
since (or until) the orbit would cross the barycenter of the system; 
this procedure makes the transit spacing and duration nearly uniform if 
the planet orbits with a nearly circular orbit.  
This algorithm detects Kepler-38b strongly 
with S/N $\approx 17$, 
well above the $3\sigma$ false-alarm probability of 
$\approx 10$.  
We removed the signal of Kepler-38b, re-applied the algorithm, and no 
other transiting bodies
were detected with periods between 50 to 300 days with 
sizes larger than 
$\approx 0.7 \ R_{\rm Nep}$ $(P/105)^{1/2}$ 
at $>3\sigma$ significance, 
where $R_{\rm Nep}$ is the radius of Neptune and
$P$ is the orbital period in days.

In Kepler-16, Kepler-34, and Kepler-35 the circumbinary planets
gravitationally perturb the stellar orbits, leading to phase changes
of the secondary eclipse relative to the primary eclipse
\citep{Doyle_2011,Welsh_2012}.  Such phase changes can be
characterized by measuring accurate eclipse times for all primary and
secondary eclipses, fitting a linear ephemeris using a common period
and independent zero-points, and plotting the residuals on a Common
Period Observed minus Computed (CPOC) diagram.  In a case such as
Kepler-34 where the gravitational perturbation from 
the planet is relatively strong,
the residuals of the primary eclipse times have a different slope than
the residuals of the secondary eclipses in the CPOC diagram
\citep{Welsh_2012}.  Over long time-scales, the phase changes become
cyclic, and the signals in the CPOC diagram are poorly described by
linear fits.  However, if the time-scale of the observations is short
compared to the precessional time-scale, then the CPOC signals of the
primary and secondary will be approximately linear and fitting a
linear ephemeris to the primary eclipse times and to the secondary
eclipse times separately will yield different periods.

For Kepler-38, the times for the primary and secondary eclipses were
measured by fitting the eclipse with a low-order cubic Hermite
polynomial, as described in 
\citet{Steffen_2011} and
\citet{Welsh_2012}.
The typical uncertainties are about 30
seconds for the primary eclipses and about 9 minutes for the secondary
eclipses (see Table 2).  Linear fits to the primary and secondary
eclipse times were performed separately.  The uncertainties on the
individual times were scaled to give $\chi^2\approx N$, where $N$ is
the number of primary or secondary eclipse times, and the following
ephemerides were arrived at:

\begin{tabular}{lll}
$P= 18.7952679\pm 0.0000029$\, {d}, 
    &  $T_0 = 2,454,952.872560\pm 0.000090$  & {primary}  \\
$P = 18.795224\pm 0.000051$\, {d}, 
    &  $T_0 = 2,454,962.2430\pm 0.0016$   & {secondary},  \\
\end{tabular}

\noindent where the times are barycentric Julian dates (BJD\_TDB).
The difference between these primary and secondary periods is $3.79\pm
4.40$ seconds.  The CPOC signals for the primary eclipses and the
secondary eclipses are parallel and show no periodicities, which is an
indication there are no phase changes of the secondary eclipse
relative to primary eclipse.  Thus the object in the circumbinary
orbit in Kepler-38 has essentially no observable gravitational effect
on the binary, at least on a time-scale of a few years.

\subsection{High Resolution Spectroscopy}

Kepler-38 was observed from the McDonald Observatory with
the Harlan J. Smith 2.7 m Telescope (HJST) 
and the Hobby-Eberly Telescope (HET)
between 2012 March~28 and May~1.  The HJST was equipped with
the Tull Coud\'e Spectrograph \citep{tull1995}, which covers
the entire optical spectrum at a resolving power of 
$R=60,000$. At each visit we
took three 1200-second exposures that were then
co-added, assuming no Doppler shifts between them.
The time of the observation was adjusted to the mid-time
of the exposure, properly accounting for the CCD readout
time.
The radial velocity standard star HD 182488 was usually
observed in
conjunction with Kepler-38.
A Th-Ar hollow cathode lamp was frequently observed to provide
the wavelength calibration.
In all, a total of nine HJST observations 
of Kepler-38 were obtained.
A customized IRAF script was used to reduce the data, 
including
the correction for the electronic bias, 
the scattered light subtraction,
the flat-field correction, the optimal aperture extraction, 
and the wavelength calibration.
The signal-to-noise ratio at the peak of the order
containing the Mg b feature near 5169~\AA\ ranged from
10.3 to 15.9 per pixel.
The HET was equipped with
the High Resolution Spectrograph (HRS, Tull 1998), 
and the configuration we used had 
a resolving
power of $R=30,000$ and a wavelength coverage of
about 4800~\AA\ to 6800~\AA.
During each visit to Kepler-38 we also obtained
a spectrum of HD 182488 and spectra of a Th-Ar lamp.
A total of six
observations of Kepler-38 were obtained.
The images were reduced and the spectra were extracted
in a manner similar to the HJST spectra, 
but with 
software customized for the HET+HRS configuration.
The signal-to-noise per pixel at the peak of the
order with the Mg b feature ranged from 19.4 to
45.6

We used the ``broadening function'' technique \citep{rucinski1992} 
to measure the radial velocities.  
The broadening functions (BFs) are essentially rotational
broadening kernels, where the centroid of the peak yields
the Doppler shift and where the width of the peak is
a measure of the rotational broadening.
In order to make full use of this technique, one must have
a high signal-to-noise  spectrum of a slowly rotating
template star observed with the same instrumentation as the target
spectra.  We used
observations of HD182488 (spectral type G8V)
for this purpose
for each respective data set (HJST, HET).
The radial velocity of this star
was taken to be
$-21.508$
km s$^{-1}$
\citep{Nidever_2002}.

The spectra were prepared for the BF analysis
by merging the echelle orders using a two-step process.
First, each order was normalized to its local continuum
using cubic splines.  The low signal-to-noise ends of
each order were then trimmed so that there was
only a modest wavelength
overlap ($\approx 5-10\%$) between adjacent orders.  The normalized
orders were then co-added and interpolated to a log-linear 
wavelength scale.  The wavelength range used in the analysis was
5137.51 to 5509.01~\AA\ for the HJST spectra and
4830.00 to 5769.95~\AA\ for the HET spectra.

\begin{figure}
\centerline{\includegraphics[scale=0.75,
angle=0]{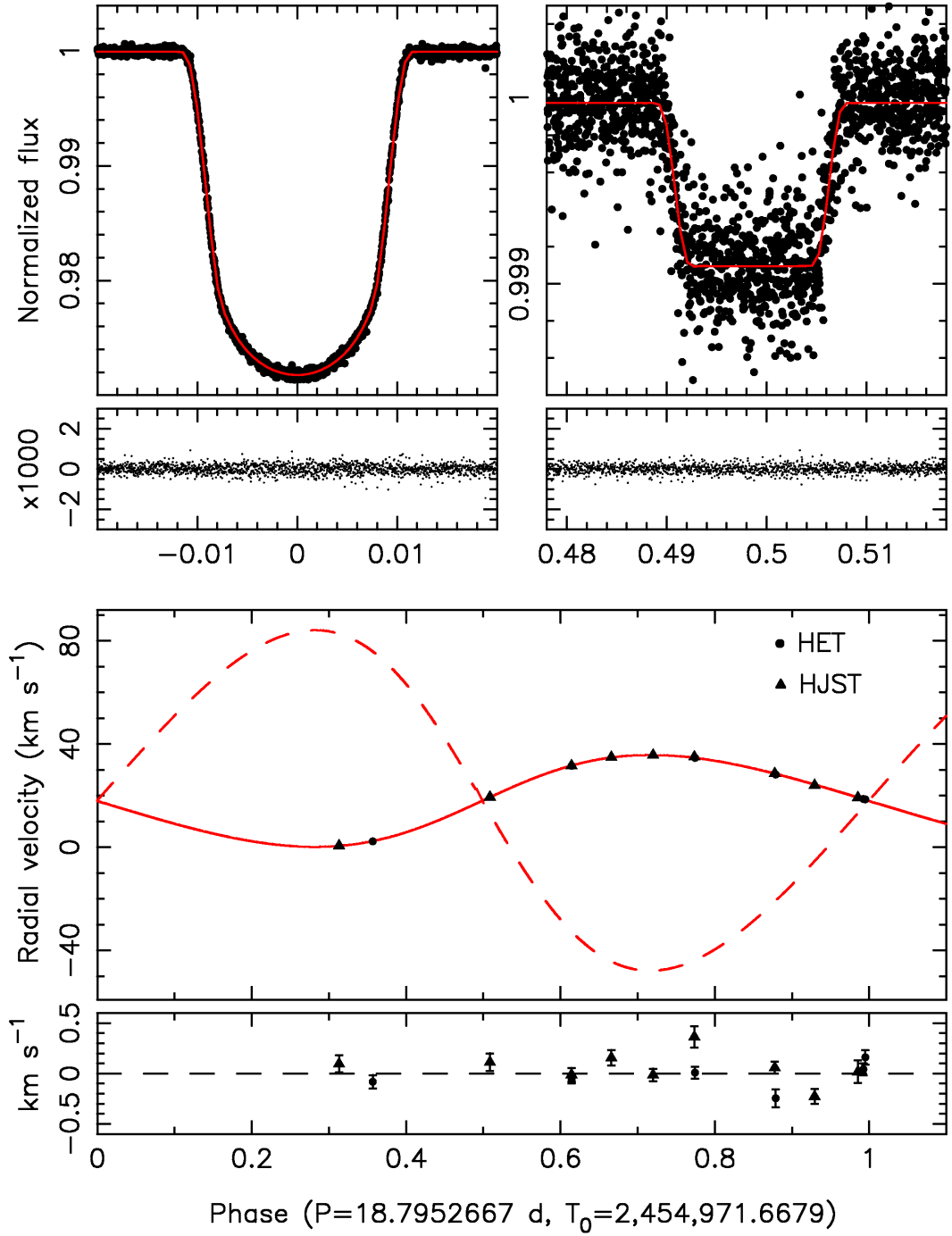}}
\caption{Top: The primary and secondary
eclipse profiles and the ELC fits.  
The standard deviation of the residuals is about 210 ppm.
Bottom: The radial velocities of the primary and the best-fitting
ELC model curve.  The dashed curve is the predicted radial
velocity curve of the secondary star.
\label{showlc}}
\end{figure}

From the BF analysis we found that
the spectra were all single-lined (i.e.\ only one
peak was evident in the BFs).  
Using some simple numerical simulations,
we estimate that
the lack of a second peak places a lower limit on the
flux ratio of the two stars of $F_A/F_B\gtrsim 25$
in the HET spectral bandpass, where the subscript $A$ refers
to the brighter primary star and $B$ refers to the fainter secondary 
star.
A Gaussian was fit to the BF peaks to determine the
peak centroids.   Barycentric corrections and a correction
for the template radial velocity were applied to the 
BF peak velocities to arrive at the
final radial velocity measurements (see Table \ref{RVs} and
Figure \ref{showlc}).

\begin{figure}
\includegraphics[scale=0.7,angle=-90]{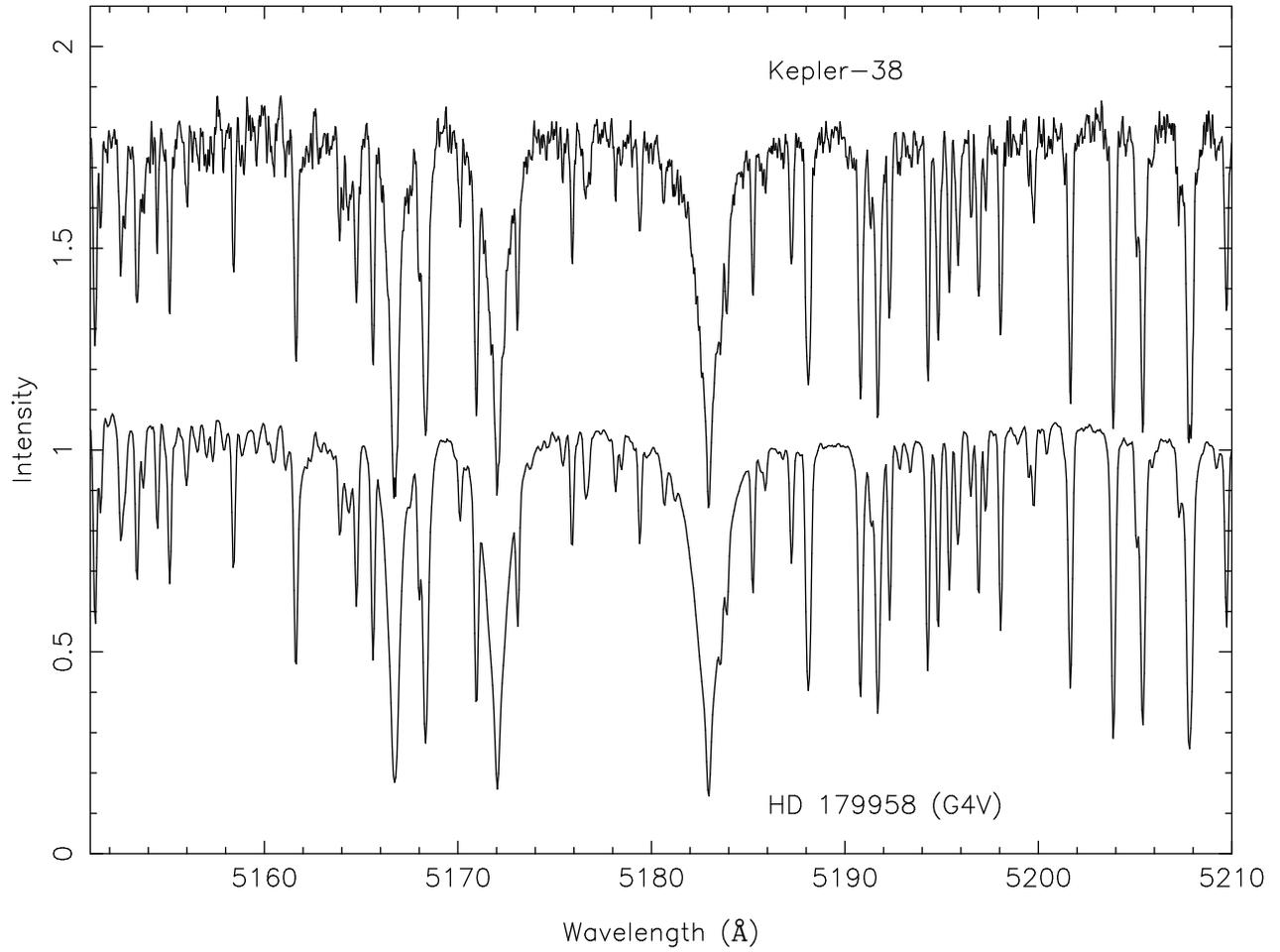}
\caption{The average ``restframe'' HET spectrum of
Kepler-38 (top, shifted vertically by 0.5 units)
and the HET
spectrum of the G4V star HD 179958.
Overall the match is quite good.
\label{plotspec}}
\end{figure}

The six HET spectra were Doppler corrected to zero velocity and co-added
to create a ``restframe'' spectrum covering the range 4830-6800~\AA.
We cross-correlated this spectrum against a library of 
F-, G- and K-type dwarfs obtained with the HET/HRS (but with a different
cross-disperser configuration).  The spectrum of the G4V star
HD 179958 provides a good match as shown in Figure \ref{plotspec}.

The signal-to-noise of the spectra proved to be too low
to attempt a determination of the
stellar temperature, gravity, and metallicity via the
measurements of individual lines.  Instead, 
we used the Stellar Parameter
Classification (SPC) code \citep{Buchhave_2012} to measure
the spectroscopic parameters.  The SPC analysis is well suited
for spectra with relatively low signal-to-noise.  The observed
stellar spectrum is cross-correlated against a dense
grid of synthetic spectra  that
consists of 51,359 models covering
a wide range in effective temperatures, gravities,
metallicities, and rotational velocities.  Since the SPC
analysis uses all of the absorption lines
in the wavelength region
5050 to 5360~\AA, the metallicity will be denoted as [m/H].
However, in practice we do not expect [m/H] to be significantly
different from [Fe/H] for a star that is close to solar metallicity,
and in the following we will consider them to be equivalent.
The HJST spectra were combined to yield a spectrum with a signal-to-noise
of $\approx 53$ at the peak of the echelle order that contains
the Mg~b features near 5169~\AA.   
The HET spectra were likewise
combined to yield a spectrum with a signal-to-noise
ratio of $\approx 196$ near the Mg~b feature.
Separate fits were done with the gravity as a free parameter and with
the gravity fixed at the dynamical
value of $\log g=3.926$,
since the dynamically determined
gravity is 
fairly 
robust and has a small uncertainty.
The SPC derived parameters are given in  
Table \ref{erik}.   The gravity found by SPC agrees with the dynamical
gravity at the $1\sigma$ level.  The effective temperature of
$T_{\rm eff}=5623\pm 50$~K is roughly what one expects for
a spectral type of G4V.
For the final adopted parameter
values,
we use the dynamically determined gravity, and simply average the
SPC-derived values from the HJST and HET spectra with the gravity fixed.

\section{Light and Velocity Curve Models}

\subsection{Photometric Dynamical Model}

The light and velocity curves of Kepler-38 were modeled using
the photometric dynamical model described in Carter et al.\ (2011;
see also Doyle et al.\ 2011 and Welsh et al.\ 2012 for previous
applications to transiting circumbinary planet systems).  
The code integrates the equations of motion for three bodies
and, when given a reference time and viewing angle, synthesizes the
light curve by accounting for eclipses and transits as necessary,
assuming spherical bodies.
The radial velocities of the components are also computed as a function
of time. 
Normally,
because Kepler-38 is a single-lined binary, one would have to assume
a mass for the primary or assume a mass ratio to fully solve
for the component masses and radii.  Fortunately, the presence of transits 
constrain the dynamical solution since their exact timing depends in
part on the binary mass ratio.  On the other hand, the planet
so far has had no measurable effect on the eclipse times of the primary
and secondary.  Given this, the stellar masses and
the mass of the planet are not as tightly constrained as
they were in the cases of Kepler-16, Kepler-34, and Kepler-35.

\begin{figure}
\begin{center}
\includegraphics[scale=1.1,angle=0]{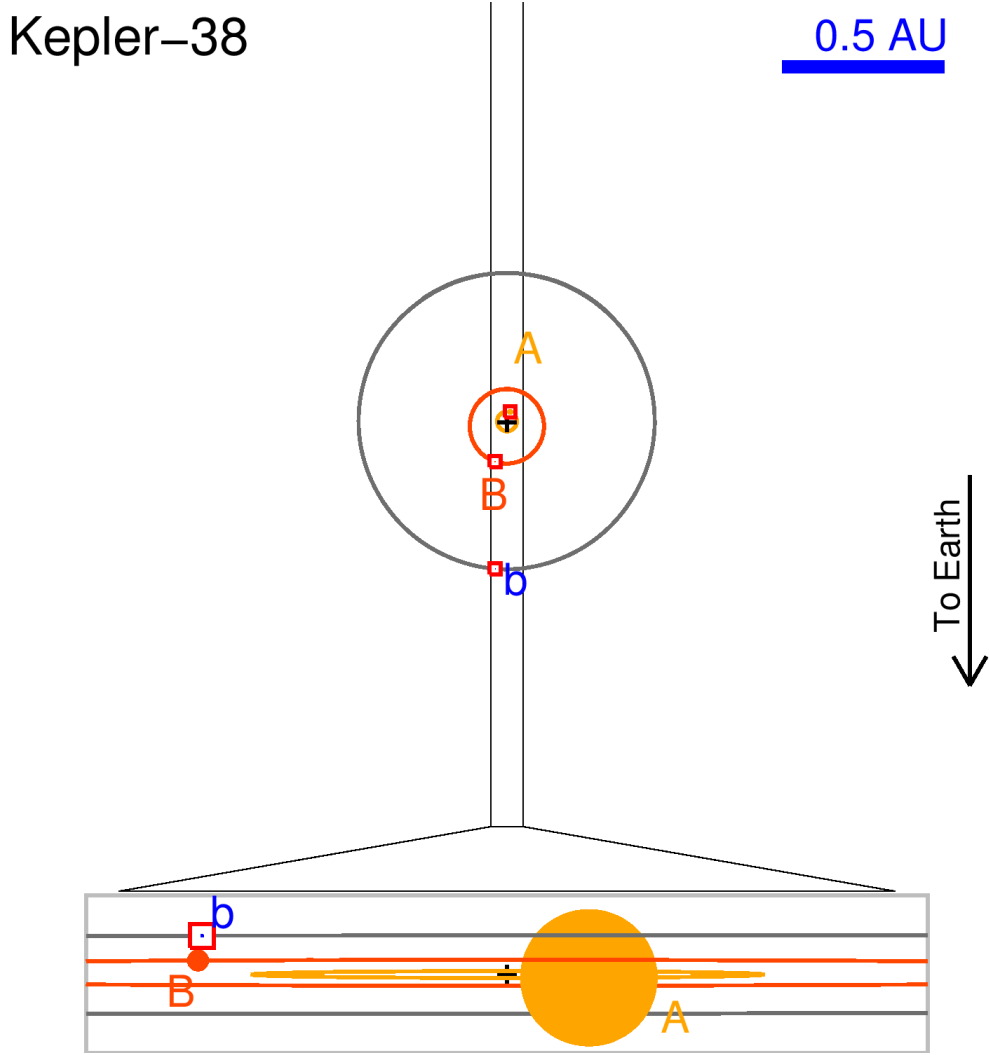}
\end{center}
\caption{Top:  A scaled, face-on view of the orbits in the Kepler-38 system
is shown.  The configuration shown is correct for the reference
epoch given in Table 1.
On this scale the stars
and the planet are too small to be seen and are represented
by the small boxes.  The labels A, B, and b denote the primary
star, the secondary star, and the planet, respectively.
Bottom:  The region between the vertical lines
in the top diagram is
shown on an expanded scale with an  
orientation  corresponding to what would be seen from Earth.
Transits
of b across A are observed, and occultations of b due to A occur but
are not observable given the noise level.  Transits of b across B and
occultations of b due to B do not occur in this configuration.
\label{schem}}
\end{figure}

The model as applied to Kepler-38 has 34 parameters, including parameters
related to the masses (the mass of Star A, the binary mass ratio, and the
planetary to binary
 mass ratio), the stellar and planetary
orbits (the period, the reference time
of primary eclipse, the inclination, and eccentricity/orientation parameters),
radius and light parameters (the fractional stellar radii, 
the planetary to primary radius
ratio, 
the stellar flux ratios, and the limb darkening parameters), relative
contaminations for each 
of the 11 Quarters of data, a light curve noise scaling 
parameter, and radial velocity zero points for the HET and HJST measurements.
The model was refined 
using a 
Monte Carlo Markov Chain routine
to estimate
the credible intervals for the model parameters.
The resulting best-fitting parameters and their uncertainties are summarized
in Table
\ref{photodynamicalparm}, and
derived astrophysical parameters of interest are summarized
in Table \ref{derived}.   
Figure \ref{koi1740_transits} shows the transits and the best-fitting
model, and Figure \ref{schem} shows  schematic diagrams of the
stellar and planetary orbits.

The circumbinary planets in Kepler-16, Kepler-34, and Kepler-35 transit
both the primary star and the
secondary star.  When transit events
across both stars are observed, the constraints on the orbital parameters
are much tighter than they are when only transits across the
primary are
seen.  In the case of Kepler-38, the planet does not transit
the secondary in the best-fitting model (Figure \ref{schem}),
although  
given
the uncertainty in the nodal angle $\Omega$, transits of the
secondary might occur for some of the acceptable models derived
from the Monte Carlo Markov Chain.  However,
individual transits of the planet across the secondary
won't be observable, owing to the extreme flux ratio
of star B to star A, where $F_{B}/ F_{A}=9\times 10^{-4}$ 
in the {\it Kepler} bandpass
(Table \ref{photodynamicalparm}).  
The expected depth of the transit
of the planet across the secondary is on the order of 21 ppm,
which is a factor of 10 smaller than the noise level of $\approx 210$ ppm.
In a similar vein, occultations of the planet by star A do occur, but
are undetectable given the noise level.

\subsection{ELC Model}

As an independent check on the parameters of the binary, we modeled
the light and velocity curves using the 
Eclipsing Light Curve
(ELC) code (Orosz \& Hauschildt
2000) with 
its genetic algorithm and Monte Carlo Markov Chain optimizers.
As noted above, we cannot use ELC to solve for the stellar masses or
the absolute stellar radii since Kepler-38 is a single-lined
binary.  However, ELC can be used to find the orbital parameters
($K$, $e$, $\omega$, $i$, $P$, and $T_{\rm conj}$), 
the fractional radii $R_A/a$ and $R_B/a$,
the temperature ratio $T_B/T_A$, and the stellar limb darkening
parameters $x_A$ and $y_A$ for the  quadratic
limb darkening law [$I(\mu)=I_0(1-x(1-\mu)-y(1-\mu)^2)$].
In the limiting case where the 
stars are sufficiently separated that their shapes are spherical,
ELC has a fast ``analytic'' mode where the equations given
in Gimenez (2006) are used.
Table \ref{ELCparm} gives the resulting parameters of the fit.
The agreement between ELC and the photometric
dynamical
model is good.  

The photometric
dynamical model assumes the stars are spherical.  We computed a model
light curve using ELC, assuming ``Roche'' geometry (to the extent 
that is possible
in an eccentric orbit, see Avni 1976 and Wilson 1979).  
At periastron, the ``point'' radius of the
primary along the line of centers
differs from the polar radius by 0.021\%.  Thus the assumption of 
spherical stars
is a good one.
The expected amplitude of the modulation of
the out-of-eclipse part of the
light curve due to reflection and ellipsoidal modulation
is $\approx 180$
ppm.  If Doppler boosting 
\citep{Loeb_2003,Zucker_2007} is included,
the amplitude
of the combined  signal from all effects
is 
$\approx 270$ ppm, with the maximum observed at phase $\phi\approx 0.3$.
Finally,
the amplitude of the
signal in the radial velocity curve due to the Rossiter-McLaughlin effect 
during the primary eclipse (when the secondary star transits the
primary) is 
on the order of 55 m s$^{-1}$ for a projected
rotational velocity of
$2.4$ km s$^{-1}$.

\section{Discussion}

\subsection{Planetary Parameters}

The radius of Kepler-38b is $4.35\,R_{\oplus}$ 
($=1.12\,R_{\rm Nep}$ or
$0.39\,R_{\rm Jup}$, using the equatorial radii),
with an uncertainty of $\pm 0.11\,R_{\oplus}$ (or 2.5\%).  
For comparison, its radius
is about half of the radius of
Kepler-16b
($R=0.7538 \pm 0.0025\,R_{\rm Jup}$, Doyle et al.\
2011), Kepler-34b ($R=0.764\pm0.014\,R_{\rm Jup}$, Welsh et al.\
2012), and Kepler 35b  ($R=0.728\pm 0.014\,R_{\rm Jup}$,
Welsh et al.\ 2012).  Thus all four of the transiting circumbinary
planets discovered
so far have radii substantially smaller than Jupiter's.  Since a 
Jupiter-sized planet would have a deeper transit and would therefore be easier
to find (all other conditions being equal), the tendency for the
circumbinary planets to be sub-Jupiter size is noteworthy.  
\citet{Pierens_2008} argued that Jupiter-{\em mass} circumbinary planets
in orbits relatively close to the binary should be rare owing to the various
instabilities that occur during the migration phase and also to
subsequent resonant
interactions with the binary (Jupiter-mass circumbinary planets 
that orbit further out
from the binary could be stable, but these would be less 
likely to transit owing to a larger separation).  
The predictions of 
\citet{Pierens_2008} seem to be consistent with what is known
from the first four {\em Kepler} transiting circumbinary planets.

Since Kepler-38b has not yet
noticeably perturbed the stellar orbits, we have only an
upper limit on its mass of $M_b<122\,M_{\oplus}$ ($<7.11\,M_{\rm Nep}$ or
$<0.384\,M_{\rm Jup}$)
at 95\% confidence.  While this is clearly a substellar mass, 
this upper limit
is not particularly constraining in terms of the density, as
we  find $\rho_{\rm b}<8.18$ g cm$^{-3}$.  
A reasonable mass is $M_b\approx 21\,M_{\oplus}$, assuming the planet 
follows the
empirical mass-radius relation of $M_b=(R_b/R_{\oplus})^{2.06}\,M_{\oplus}$
(Lissauer et al.\ 2011).

\begin{figure}
\centerline{\includegraphics[scale=0.7,angle=-90]{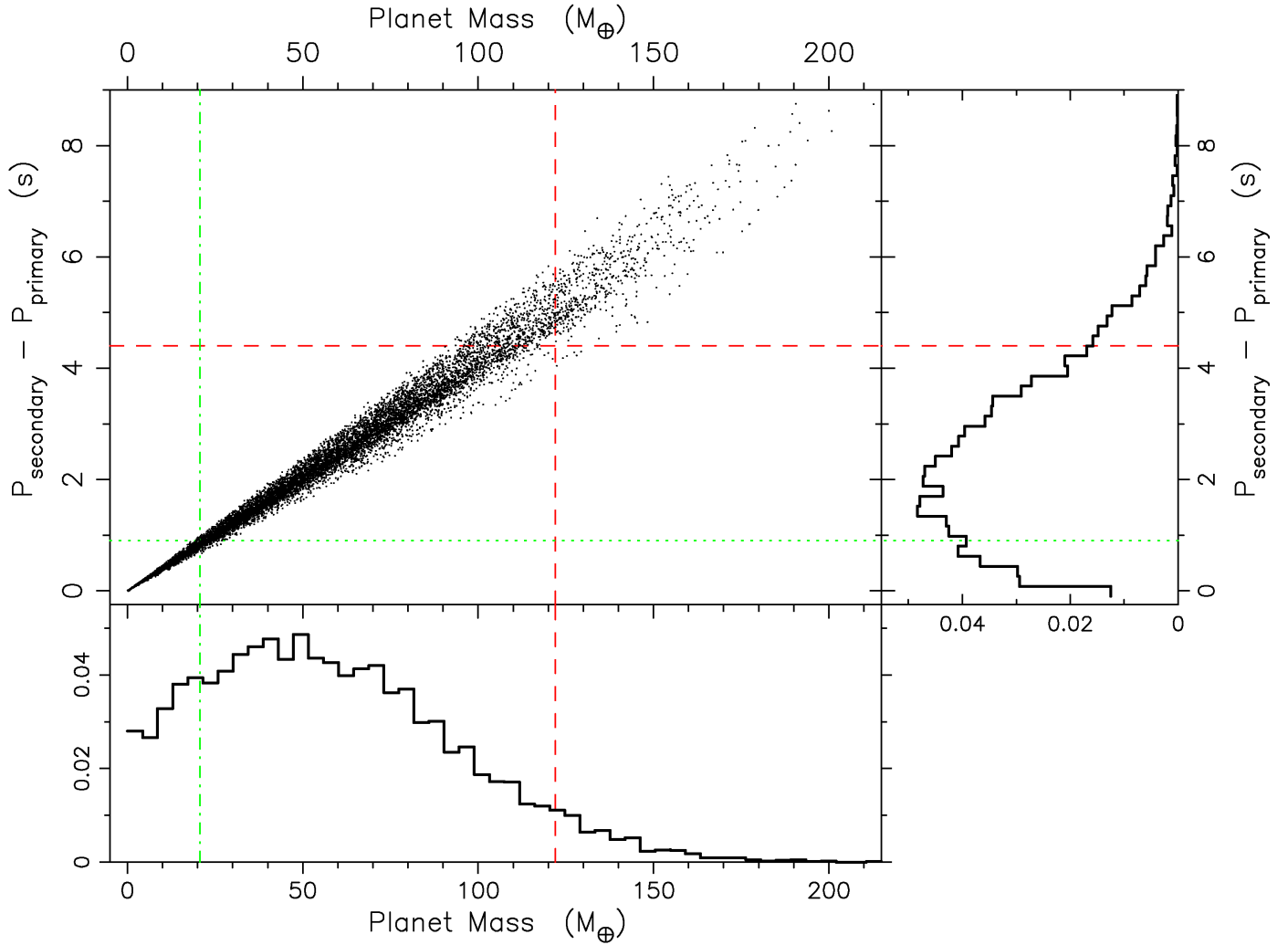}}
\caption{
A set of 10,000 acceptable solutions from the photometric dynamical model 
Markov Chain Monte Carlo is plotted, showing the differences in 
the primary and secondary eclipse periods 
versus the planetary mass.
Histograms of the distribution collapsed over mass and period difference
are also shown, where the vertical axes represent the fraction of
trials in a given bin. 
To the left of the vertical dashed line at 
$122\,M_{\oplus}$ is where 95\% of the acceptable solutions 
are situated.
The vertical dash-dotted line marks a mass of
$21\,M_{\oplus}$, which is the mass the planet would have if it
follows an empirical mass-radius relation.
The red horizontal dashed line is the current
$1\sigma$ limit
of 4.4 seconds for the difference between the primary and secondary
eclipse periods.
The green horizontal dotted line at 0.9 seconds marks the
expected location of the observational limit on the difference
between the eclipse periods that should be obtainable by the end of
the Extended Mission. 
\label{dp_vs_M}}
\end{figure}

The gravitational interaction between the planet and the two stars
causes small perturbations that will grow over time and
will eventually
lead to a measurable change in the phase difference between the primary
and secondary eclipses.  As discussed above, this change in the
phase difference manifests itself as a
difference between the period measured from the
primary eclipses and the period measured
from the secondary eclipses.
In Kepler-16, Kepler-34, and Kepler-35, the time-scale for a measurable
period difference to occur is relatively short, as divergent periods were
measured using 6 Quarters of data for Kepler-16, and 9 Quarters of data
for Kepler-34 and Kepler-35.  For Kepler-38, 
the $1\sigma$ limit on the period difference is $<4.4$ seconds using
11 Quarters.
Figure \ref{dp_vs_M} shows the set of 
acceptable planetary masses and the changes in orbital period 
the planet would induce, based on the Monte Carlo Markov Chain
from the photometric dynamical model. 
The larger the planetary mass,
the larger the difference in periods it causes between the primary and 
secondary star.
The lack of any measurable period difference places an upper limit
on the mass of the planet: 
To the left of the vertical dashed line at 122 $M_{\oplus}$ 
is where 95\% of the acceptable solutions reside.
The vertical dot-dash line marks a planetary mass of 21 $M_{\oplus}$.
The horizontal dashed line at 4.4 seconds 
marks the observed $1\sigma$ uncertainty 
in the measured value of $P_{2} - P_{1}$ accumulated over the span of 
the current {\it Kepler} observations (52 binary star eclipses); 
valid planetary masses lie below this line, roughly.
The dotted horizontal line at 0.9 seconds
marks a $1\sigma$ period difference 
uncertainty that can be placed when $\approx 150$ eclipses are eventually 
observed by {\it Kepler}, corresponding approximately to the end of the 
Extended Mission in the year 2017.
If no period difference is measured at that time, 
the mass of the planet would be less than $\approx  20\,M_{\oplus}$
with $1\sigma$ uncertainty.
(The intersection of this period difference {\em uncertainty} and 
the 21 $M_{\oplus}$  line within the set of Monte Carlo points 
is a coincidence.)  
If the planet has a normal density and a mass of $21\,M_{\oplus}$,
then the period difference would be $\approx 0.9$ seconds, which
would require $\approx 312$ binary orbits or about 16 years 
to obtain a $3\sigma$ detection.  

\subsection{Stellar Parameters and Age}

\begin{figure}
\centerline{\includegraphics[scale=0.86,angle=0]{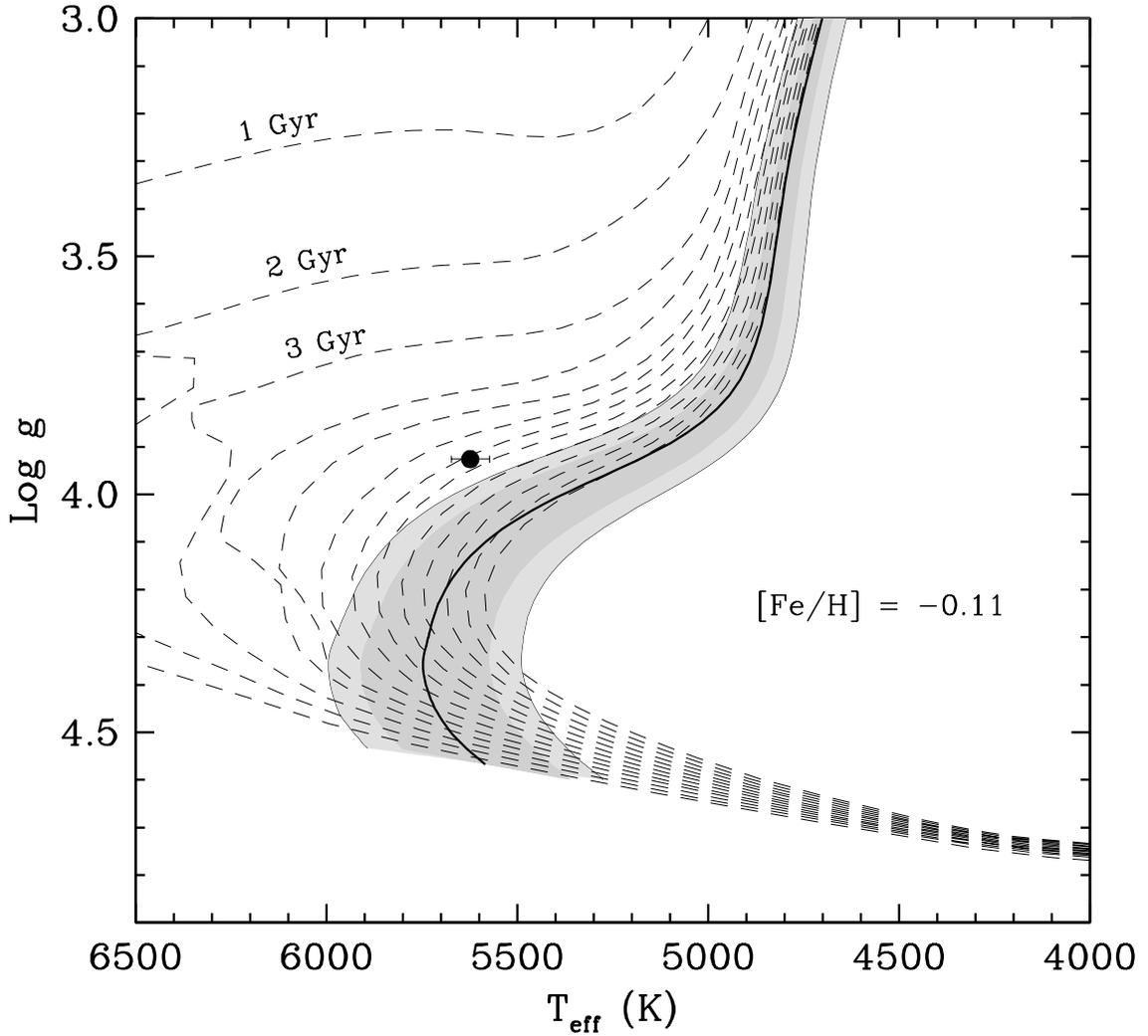}}
\caption{A $\log g-T_{\rm eff}$ diagram
showing the location of the primary of
Kepler-38 as determined from the spectroscopic
analysis and the photometric
dynamical model (black point).  The solid
line is an Yale evolutionary track appropriate for the
measured mass
and spectroscopic metallicity
(assuming equivalency
between [m/H] and [Fe/H]). 
The heavy shaded region shows the $1\sigma$ error
region accounting for the uncertainty in the mass,
and the lighter shaded region shows the $1\sigma$ uncertainty
region when the uncertainty in the metallicity is also included.
The dashed lines represent
isochrones with the same metallicity and ages of 1--13 Gyr (left to
right).
\label{iso}}
\end{figure}

\begin{figure}
\centerline{\includegraphics[scale=0.8,angle=0]{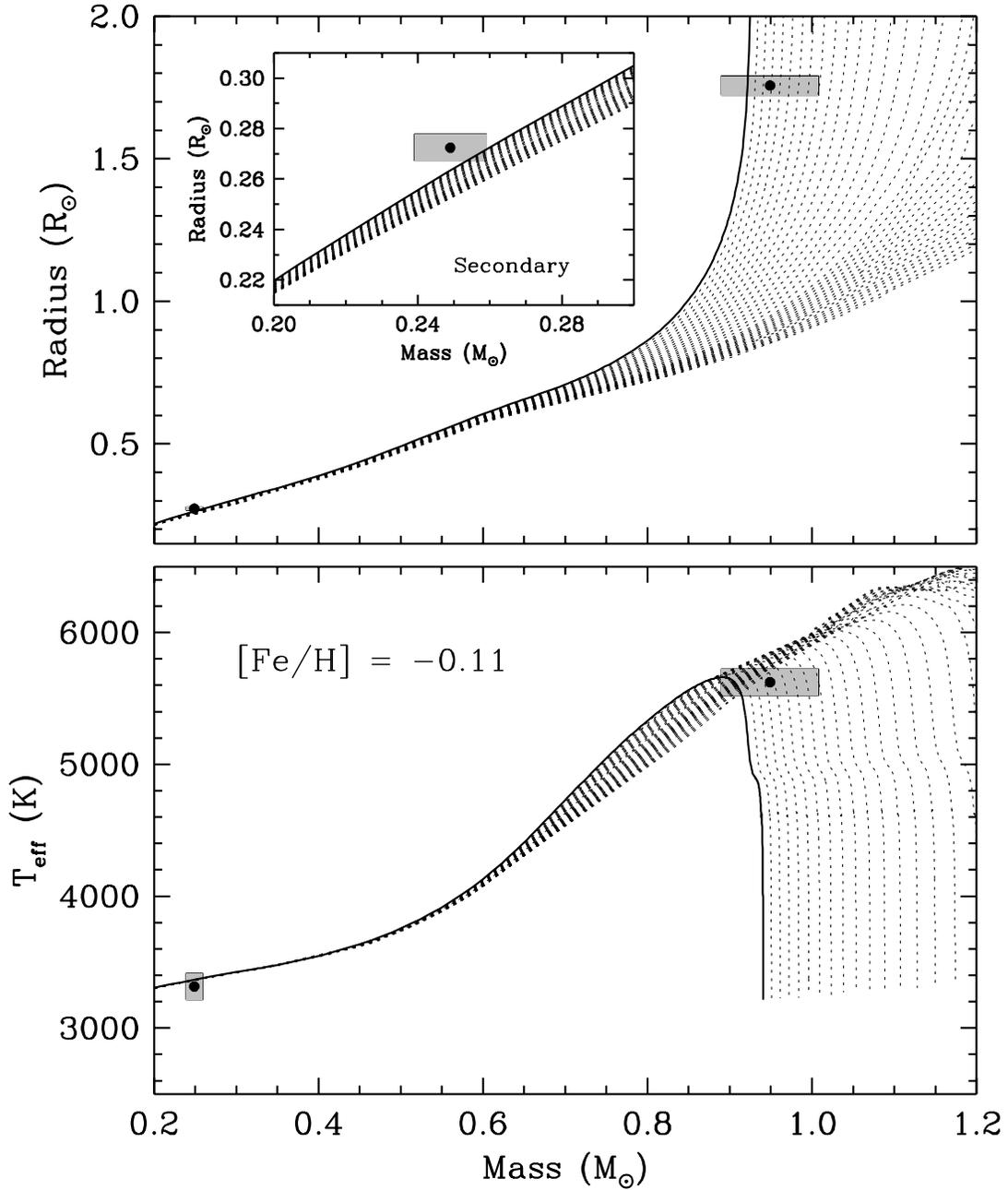}}
\caption{
Mass-radius and mass-temperature diagrams showing the locations of
the primary and secondary stars in Kepler-38. The dotted lines
correspond to model isochrones from the Dartmouth series (Dotter et
al.\ 2008) for the measured metallicity (assuming equivalency between
[m/H] and [Fe/H]) and ages from 1 to 13 Gyr. The oldest isochrone is
represented with a heavy solid line.
\label{massrad}}
\end{figure}

As noted earlier, the observed timing of the planet transits
and the amplitude of the radial velocity curve sets the scale
of the binary, and we are able to measure masses and radii for each
star.  Considering the fact that Kepler-38 is a single-lined spectroscopic
binary (note that the ratio of the secondary-to-primary
flux in the {\em Kepler} bandpass is $9\times 10^{-4}$, see
Table \ref{photodynamicalparm}),
the uncertainties in the masses and
radii are fairly small
(6.2\% 
and 3.6\% for the primary and secondary  masses, respectively, and
1.8\% for the radii).   On the other hand, these uncertainties
are still somewhat larger than what one would like when doing
precise comparisons with stellar evolutionary models
(for comparison, the
stellar masses and
radii are known to much better than 1\% for the first 
three {\em Kepler} circumbinary planets, see Doyle et al.\ 2011
and Welsh et al.\ 2012). 
Clearly, since
the radius of the primary star ($1.757\pm 0.034\,R_{\odot}$)
is much larger
than the expected zero-age
main sequence radius for its mass
($0.949\pm 0.059\,M_{\odot}$), the primary must be significantly evolved,
but is still a core hydrogen-burning star.
The situation is
shown in Figure \ref{iso}, which gives the position of Kepler-38
in a $T_{\rm eff}-\log g$ diagram. The heavy solid line is a
Yonsei-Yale evolutionary track (Yi et al.\ 2001)
for $[{\rm Fe/H}]=-0.11$, which is our adopted spectroscopic
metallicity determination (Table 4). 
 As indicated earlier, we assume
here that the iron abundance is well approximated by the [m/H]
metallicity index measured with SPC. 
The dark shaded area is
the uncertainty in the location of the track that results from
the uncertainty in the mass, and the lighter shaded area also
includes the uncertainty in the metallicity.  The observed 
$T_{\rm eff}$ and $\log g$
is just outside the $1\sigma$ region of the evolutionary track.
Figure \ref{massrad} shows the positions of 
the stars in Kepler-38 on 
the mass-radius and mass-temperature diagrams
[the temperature of the secondary was inferred from
the measured temperature of the primary 
(Table \ref{erik})
and the temperature
ratio from the ELC models (Table \ref{ELCparm})].
They are compared against
model isochrones from the Dartmouth series (Dotter et al. 2008), in
which the physical ingredients such as the equation of state and the
boundary conditions are designed to better approximate low-mass stars
(see, e.g., Feiden et al.\ 2011). 
If we use the mass, radius, and metallicity of the primary, the 
inferred age is 13 Gyr.  If, on the other hand, we use the measured
temperature, the age of the system would be between 7 and 8 Gyr.
Accounting for the small discrepancy between the measured temperature
and the other parameters, we adopt an age of $10\pm 3$ Gyr.

The secondary star has a relatively low mass
($0.249\pm 0.010\,M_{\odot}$), and 
is one of just a handful of
low mass stars with a well-measured mass and radius.
Its mass is slightly 
larger than those of CM~Dra~A 
($0.2310\, M_{\odot}$; Morales et al.\ 2009)
and KOI-126~B 
($0.20133\,M_{\odot}$; Carter, et al.\ 2011),
and Kepler-16b ($0.20255\pm 0.00066\,M_{\odot}$;
Doyle et al.\ 2011,  
see also Bender et al.\ 2012 and Winn et al.\ 2011).
Stars whose mass is $<0.8\,M_{\odot}$ typically
have radii that are $\sim 10-15\%$ larger 
than what is predicted by stellar evolutionary models 
(Torres \& Ribas 2002; Ribas 2006; Ribas et al.\ 2006;
L\'opez-Morales 2007;
Torres et al.\ 2010; Feiden et al.\ 2011).   Relatively high levels of
stellar activity induced by tidal interactions in short-period
systems is one possible cause of this discrepancy (L\'opez-Morales
2007).  There is a hint that the secondary star is inflated, although we note
there is a small discrepancy
with the models for the primary.

\subsection{Stellar Variability}

With {\it Kepler} data, it is frequently possible to determine the rotation 
period of the star from the modulations in the light curve due to star-spots.
In the case of Kepler-38 however, the modulations are small (rms $<600$ 
ppm for the long cadence time series, omitting the eclipses), and 
considerably smaller than the instrumental systematic trends in the light 
curve. Separating intrinsic stellar variability from instrumental 
artifacts is difficult, so we used the PDC 
light curve that has many of
the instrumental trends removed.
The PDC data for Kepler-38
are generally flat outside of eclipses, with the exception of
Quarter 1, which we therefore omitted. The data were normalized and
the primary and secondary eclipses removed from the time series.
A power spectrum/periodogram was computed, and the dominant 
frequency present corresponds to the orbital period (18.79 days). 
We also patched the gaps in the light curve with a random walk and 
computed the auto-correlation function (ACF). The ACF revealed a 
broad peak at $\sim 18$ days, consistent with the orbital period.

The fact that the orbital period is manifest in the out-of-eclipse 
light curve initially suggested that star-spots were present and the
star's spin was tidally locked with the orbital period. But the
non-zero eccentricity of the orbit means that exact synchronicity is
impossible (see below). Phase-folding the data on the orbital period,
and then binning to reduce the noise, revealed the origin of the orbital 
modulation: Doppler boosting combined with reflection and ellipsoidal
modulations. 
The amplitude is 
$\sim 300$ ppm with a maximum near phase $\phi=0.25$,
consistent
with what is expected from the ELC models.
We verified that the Doppler boost signal is also present in the
SAP light curves.
For more discussion of Doppler boosting in {\em Kepler}
light curves, 
see
Faigler \& Mazeh (2011) and
Shporer et al.\ (2011).

For an eccentric orbit, there is no one spin period that can be 
synchronous over the entire orbit. However, there is a spin period
such that, integrated over an orbit, there is no {\em net} torque on 
the star's spin caused by the companion star.
At this ``pseudosynchronous'' period, the spin is in equilibrium and
will not evolve (Hut 1981, 1982).  
Given the old age of the binary, it
should have reached this pseudosynchronous equilibrium state.
The ratio of orbital period to
pseudosynchronous spin period is a function of the eccentricity only 
(Hut 1981), and for Kepler-38 this period is $P_{\rm pseudo}=17.7$ days.
This is close to, but slightly shorter than, the $\sim 18.79$ day 
orbital period.
Using this pseudosynchronous spin period and the measured stellar 
radius, a projected rotational velocity of
$V_{\rm rot} \sin{i}\approx 4.7$ km s$^{-1}$ 
is expected, if the star's spin axis is aligned with the
binary orbital axis. 
Our spectral modeling yields $V_{\rm rot} \sin{i} = 2.4
\pm 0.5$ km s$^{-1}$, which is close to 
the rotational velocity expected due
to pseudosynchronous rotation.
Given that the rotational velocity is near the
spectral resolution limit, we can't rule out systematic errors
of a few km s$^{-1}$ caused by changes in the
instrumental point-spread function, macroturbulence, etc., 
which could bring the measured value
of $V_{\rm rot}\sin i$ up to the pseudosynchronous value.

\subsection{Orbital Stability and the Habitable Zone}

The 105 day planetary orbital period is the shortest among
the first four {\em Kepler} circumbinary planets.
The planet orbits the binary quite closely: 
the ratio of planetary to stellar orbital periods is 5.6 
(ratio of semi-major 
axes is 3.2). 
Such a tight orbit is subject to dynamical perturbations, and 
following the analytic approximation given by Holman \& Wiegert (1999), the 
critical orbital period in this binary below which the planet's orbit could 
experience an instability is 81 days. 
This compares favorably with the results of direct $N$-body
integrations, which yield a critical orbit period of 75 days.
Thus while stable, the planet is only
42\% above the critical period (or 26\% beyond the critical semi-major axis).
Kepler-38 thus joins 
Kepler-16 (14\%), 
Kepler-34 (21\%), and 
Kepler-35 (24\%) 
as systems where the planet's
orbital period is only modestly larger than the threshold for stability.
The fact that the first four circumbinary planets detected by 
{\em Kepler} are
close to the inner stability limit is
an interesting orbital feature that may be explained by processes such as
planetary migration and planet-planet
scattering during and/or post formation of these objects.
For example,
it is also possible that
strong planet migration will  bring planets
close in: migration may cease near the instability separation leading to
a pile-up just outside the critical radius; or planets that continue to 
migrate in are dynamically ejected, leaving only those outside the critical
radius \citep{Pierens_2008}.
There is also
an observational bias since objects that orbit closer to their
host star(s) will be more likely to transit, making them more likely
to be discovered.

Regardless of the cause, the close-to-critical
orbits have an interesting consequence. 
Many {\em Kepler} eclipsing binaries have orbital periods in the range
15-50 days and have
G and K type stars \citep{Prsa_2011,Slawson_2011},
and for such binaries the critical separation (roughly 2-4 times
the binary separation, depending on the eccentricity of the binary)
is close to the habitable zone{\footnote{For a 
binary star system, the habitable 
zone is no longer a spherical shell but a more complex shape
that rotates with the binary.}}.
Thus, observed circumbinary planets may preferentially lie close 
to their habitable zones. Kepler-16b is just slightly exterior to its 
habitable zone, while Kepler-34b is slightly interior (too hot).
Kepler-38b is well interior to its habitable zone, with a mean equilibrium 
temperature of $T_{\rm eq}=475$ K, assuming a Bond albedo of 0.34 
(similar to that of Jupiter and Saturn). Although $T_{\rm eq}$ is 
somewhat
insensitive to the albedo, this temperature estimation neglects the
atmosphere of the planet, and therefore should be considered a lower limit.
It is interesting to consider the situation at
a much earlier time in the past when the primary
was near the zero-age main sequence. 
The primary's luminosity would have a factor $\approx 3$ smaller,
and the equilibrium temperature
of the planet would have been $T_{\rm eq}\approx 361$~K, assuming
a similar orbit and Bond albedo.

\section{Summary}

Kepler-38b is the fourth circumbinary planet discovered by
{\em Kepler}.  The planet orbits in a nearly circular
105 day orbit about an 18.8 day, single-lined eclipsing
binary.
Using the transits and eclipses in the light curves along
with the radial velocity curve of the primary allows us
to solve for the masses and radii of the two stars 
($M_A=0.949\pm 0.059\,M_{\odot}$, $R_A=1.757\pm 0.034\,R_{\odot}$,
$M_B=0.249\pm 0.010\,M_{\odot}$, and $R_B=0.02724\pm 0.0053\,R_{\odot}$)
and
the radius of the planet
($R_b=4.35\,R_{\oplus}$). Since the gravitational interaction
between the planet and the two stars is small, we are only
able to place an upper limit of $M_b<122\,M_{\oplus}$ on
the mass of the planet.  
The first four {\em Kepler} circumbinary
planets all have a tendency to have radii substantially smaller than
Jupiter's and to have orbits that are only modestly larger than the threshold
for stability.  These tendencies yield clues into the formation, migration, and
subsequent evolution of circumbinary planets
[e.g.\ \citet{Meschiari_2012} and \citet{Paardekooper_2012}].

\acknowledgments

Kepler was selected as the 10th mission of the Discovery Program. Funding 
for this mission is provided by NASA, Science Mission Directorate. 
JAO and WFW acknowledge support from the Kepler Participating Scientist 
Program via NASA grant NNX12AD23G.  Support was also provided by the
National Science Foundation via grants  AST-1109928 to JAO, WFW, and GW,
AST-0908642 to RW, AST-0645416 to EA, and AST-1007992 to GT.

\clearpage

\begin{deluxetable}{rrrrrr}
\tablecaption{Times of Planet Transit Across the Primary\label{transittimes}}
\tablewidth{0pt}
\tablehead{
\colhead{N} &          
\colhead{Time}  &
\colhead{Uncertainty} &
\colhead{Width} &
\colhead{Depth} &
\colhead{Binary phase} \\
\colhead{} &          
\colhead{(BJD - 2,455,000)}  &
\colhead{(minutes)} &
\colhead{(days)} &
\colhead{ }  &
\colhead{}
}
\startdata
1 &  36.10896  & 14  &  0.825 & 0.0005 & 0.43 \\
2\tablenotemark{a} & 140.45975  & \nodata   & \nodata   & \nodata   & 0.98   \\
3 & 244.00847  & 26  &  0.800 & 0.0005 & 0.49 \\
4\tablenotemark{a} & 347.73815  & \nodata   & \nodata   & \nodata   & 0.02   \\
5 & 451.93756  & 18  &  0.884 & 0.0005 & 0.55 \\
6\tablenotemark{b} & 555.0~~~~   & \nodata   & \nodata   & \nodata  & 0.04 \\
7 & 659.71862  & 18  &  0.701 & 0.0004 & 0.61 \\
8 & 762.33371  & 3   &  0.348 & 0.0004 & 0.07 \\
9 & 867.38194  & 11  &  0.600 & 0.0005 & 0.66 \\
\enddata
\tablenotetext{a}{Transit is blended with primary eclipse.}
\tablenotetext{b}{Transit is in data gap.}
\end{deluxetable}

\clearpage

\begin{deluxetable}{rrrrrr}
\tablecaption{Times of Primary and Secondary Eclipses\label{eclipsetimes}}
\tablewidth{0pt}
\tablehead{
\colhead{Orbital} &          
\colhead{Primary time}  &
\colhead{Uncertainty} &
\colhead{Orbital} &
\colhead{Secondary time} &
\colhead{Uncertainty} \\
\colhead{Cycle} &          
\colhead{(BJD - 2,455,000)}  &
\colhead{(minutes)} &
\colhead{cycle} &
\colhead{(BJD - 2,455,000)}  &
\colhead{(minutes)}
}
\startdata
    1.0     &        -28.33230     &       0.420  &       1.498488  &              -18.97115  &         9.303 \\ 
    2.0     &         -9.53647     &       0.407  &       2.498488  &                \nodata  &       \nodata \\
    3.0     &          9.25767     &       0.420  &       3.498488  &              18.633540  &        8.6148 \\
    4.0     &         28.05360     &       0.434  &       4.498488  &              37.431880  &       10.3315 \\
    5.0     &         46.84874     &       0.434  &       5.498488  &                \nodata  &       \nodata \\  
    6.0     &         65.64403     &       0.420  &       6.498488  &               75.01589  &        10.218 \\ 
    7.0     &         84.43935     &       0.407  &       7.498488  &                \nodata  &       \nodata \\
    8.0     &        103.23430     &       0.476  &       8.498488  &              112.59631  &        9.4157 \\
    9.0     &        122.03006     &       0.420  &       9.498488  &              131.39748  &        10.789 \\ 
   10.0     &        140.82574     &       0.434  &      10.498488  &              150.18989  &        11.383 \\ 
   11.0     &        159.62029     &       0.420  &      11.498488  &              168.99328  &        10.102 \\ 
   12.0     &        178.41584     &       0.434  &      12.498488  &              187.78337  &         9.187 \\ 
   13.0     &        197.21093     &       0.420  &      13.498488  &              206.58514  &         8.843 \\ 
   14.0     &        216.00626     &       0.407  &      14.498488  &              225.37215  &        10.216 \\ 
   15.0     &        234.80169     &       0.420  &      15.498488  &              244.16775  &        11.023 \\ 
   16.0     &        253.59677     &       0.420  &      16.498488  &              262.96944  &        10.102 \\ 
   17.0     &        272.39233     &       0.448  &      17.498488  &              281.75459  &         9.415 \\ 
   18.0     &        291.18717     &       0.448  &      18.498488  &              300.55308  &         9.650 \\ 
   19.0     &        309.98266     &       0.448  &      19.498488  &              319.35675  &        11.936 \\
   20.0     &        328.77848     &       0.434  &      20.498488  &                \nodata  &       \nodata \\
   21.0     &        347.57494     &       0.490  &      21.498488  &              356.95425  &         9.529 \\ 
   22.0     &        366.36820     &       0.434  &      22.498488  &              375.73726  &         9.303 \\ 
   23.0     &        385.16363     &       0.420  &      23.498488  &              394.54044  &        10.216 \\ 
   24.0     &        403.95923     &       0.407  &      24.498488  &              413.32565  &        10.789 \\ 
   25.0     &        422.75416     &       0.420  &      25.498488  &                \nodata  &       \nodata \\
   26.0     &        441.54956     &       0.434  &      26.498488  &              450.92278  &        10.102 \\ 
   27.0     &        460.34465     &       0.434  &      27.498488  &              469.71921  &         9.415 \\ 
   28.0     &        479.13994     &       0.434  &      28.498488  &              488.51035  &         9.873 \\ 
   29.0     &        497.93569     &       0.434  &      29.498488  &              507.29767  &        10.789 \\ 
   30.0     &        516.73073     &       0.420  &      30.498488  &              526.10182  &         9.300 \\ 
   31.0     &        535.52578     &       0.407  &      31.498488  &                \nodata  &       \nodata \\
   32.0     &          \nodata     &     \nodata  &      32.498488  &                \nodata  &       \nodata \\
   33.0     &        573.11647     &       0.532  &      33.498488  &              582.48835  &         9.529 \\ 
   34.0     &        591.91155     &       0.434  &      34.498488  &              601.28219  &         9.187 \\ 
   35.0     &        610.70662     &       0.476  &      35.498488  &              620.07486  &         9.187 \\ 
   36.0     &        629.50191     &       0.420  &      36.498488  &                \nodata  &       \nodata \\
   37.0     &        648.29726     &       0.420  &      37.498488  &              657.66824  &         9.873 \\ 
   38.0     &        667.09279     &       0.420  &      38.498488  &              676.45686  &        13.555 \\ 
   39.0     &        685.88782     &       0.420  &      39.498488  &              695.25612  &        10.448 \\ 
   40.0     &        704.68319     &       0.420  &      40.498488  &              714.05196  &        10.102 \\ 
   41.0     &        723.47857     &       0.407  &      41.498488  &              732.84697  &         9.415 \\ 
   42.0     &        742.27339     &       0.420  &      42.498488  &              751.64364  &         9.540 \\ 
   43.0     &        761.06937     &       0.420  &      43.498488  &                \nodata  &       \nodata \\
   44.0     &        779.86425     &       0.434  &      44.498488  &              789.23446  &         9.758 \\ 
   45.0     &        798.65936     &       0.420  &      45.498488  &              808.02589  &        10.102 \\ 
   46.0     &        817.45492     &       0.420  &      46.498488  &              826.81895  &        10.102 \\ 
   47.0     &        836.25047     &       0.407  &      47.498488  &              845.61182  &         9.758 \\ 
   48.0     &        855.04557     &       0.420  &      48.498488  &              864.40410  &        10.448 \\ 
   49.0     &        873.84071     &       0.420  &      49.498488  &              883.21487  &        10.107 \\ 
   50.0     &        892.63578     &       0.434  &      50.498488  &              902.00190  &         9.873 \\ 
   51.0     &        911.43140     &       0.434  &      51.498488  &              920.80370  &         8.614 \\ 
   52.0     &        930.22640     &       0.420  &      52.498488  &                \nodata  &       \nodata \\
\enddata
\end{deluxetable}

\begin{deluxetable}{ccccc}
\tablecaption{Radial velocities of Kepler-38\label{RVs}}
\tablewidth{0pt}
\tablehead{
\colhead{Date} &          
\colhead{UT Time}  &
\colhead{BJD} &
\colhead{RV$_A$} &
\colhead{Telescope/Instrument} \\
\colhead{(YYYY-MM-DD)} &          
\colhead{}  &
\colhead{($-2$,455,000)} &
\colhead{(km s$^{-1}$)} &
\colhead{}
}
\startdata
2012-03-28 & 10:43:59.63  & 1014.96598  & $19.437  \pm   0.087$ & HJST Tull \\
2012-03-30 & 10:33:16.38  & 1016.95456  & $31.573  \pm   0.067$ & HJST Tull \\
2012-03-30 & 10:46:45.18  & 1016.95524  & $31.526  \pm   0.035$ & HET HRS \\ 
2012-03-31 & 09:48:02.35  & 1017.92319  & $34.880  \pm   0.077$ & HJST Tull \\
2012-04-01 & 10:13:30.14  & 1018.94092  & $35.685  \pm   0.062$ & HJST Tull \\
2012-04-02 & 10:19:00.60  & 1019.94478  & $34.998  \pm   0.105$ & HJST Tull \\
2012-04-02 & 10:53:40.98  & 1019.96017  & $34.613  \pm   0.059$ & HET HRS \\
2012-04-04 & 09:16:57.14  & 1021.90208  & $28.551  \pm   0.059$ & HJST Tull \\
2012-04-04 & 10:04:33.79  & 1021.92614  & $28.148  \pm   0.087$ & HET HRS \\ 
2012-04-05 & 08:37:12.35  & 1022.87403  & $24.004  \pm   0.073$ & HJST Tull \\
2012-04-06 & 09:52:28.93  & 1023.92610  & $19.222  \pm   0.112$ & HJST Tull \\
2012-04-13 & 09:39:44.77  & 1030.90929  & $~2.272  \pm   0.066$ & HET HRS \\
2012-04-25 & 08:22:56.79  & 1042.85648  & $18.693  \pm   0.059$ & HET HRS \\ 
2012-04-25 & 09:29:11.00  & 1042.90247  & $18.565  \pm   0.072$ & HET HRS \\
2012-05-01 & 08:50:50.33  & 1048.88479  & $~0.568  \pm   0.085$ & HJST Tull \\
\enddata
\end{deluxetable}

\begin{deluxetable}{lrrrrr}
\tabletypesize{\small}
\tablecaption{Spectroscopic Parameters of Kepler-38 From SPC\label{erik}}
\tablewidth{0pt}
\tablehead{
\colhead{Parameter} &          
\colhead{HJST} &
\colhead{HJST} &
\colhead{HET}  &
\colhead{HET}  &
\colhead{adopted\tablenotemark{b}}
}
\startdata
$T_{\rm eff}$ (K) 
                   & $5642\pm 50$ 
                   & $5603 \pm 50$
                   & $5603 \pm 50$ 
                   & $5643 \pm 50$ 
                   &  $5623\pm 50$\\
$\log g$ (cgs)  
               & $4.02\pm 0.10$
               & 3.926\tablenotemark{a} 
               & $3.81 \pm 0.10$
               & 3.926\tablenotemark{a} 
               & $3.926\pm 0.011$  \\ 
$[{\rm m/H}]$ (dex) 
                    & $-0.10\pm 0.08$
                    & $-0.12 \pm 0.08$ 
                    & $-0.13 \pm 0.08$
                    & $-0.10 \pm 0.08$
                    & $-0.11\pm 0.08$  \\
$V_{\rm rot}\sin i$ (km s$^{-1}$)  
                                  & $2.6\pm 0.5$ 
                    & $2.6\pm 0.5$
                    & $2.3 \pm 0.5$
                    & $2.2\pm 0.5$ 
                    & $2.4\pm 0.5$ 
\enddata
\tablenotetext{a}{The gravity was fixed at the given value.}
\tablenotetext{b}{The adopted value is the average of the HJST and
HET measurements with a fixed gravity.}
\end{deluxetable}

\clearpage

\begin{deluxetable}{lcccc}
\tabletypesize{\scriptsize}
\tablecaption{Fitting 
Parameters for Photometric Dynamical Model\label{photodynamicalparm}}
\tablewidth{0pt}
\tablehead{
\colhead{Parameter} &          
\colhead{Best fit}  &
\colhead{50\%} &
\colhead{15.8\%} &
\colhead{84.2\%} 
}
\startdata
\multicolumn{5}{c}{\bf Mass parameters}  \\
Mass of Star A, $M_A$ ($M_\odot$) & $0.949$ & $0.941 $ & $-0.059$ & $+0.055$\\
Mass ratio, Star B, $M_B/M_A$ & $0.2626$ & $0.2631 $ & $-0.0056$ & $+0.0067$\\
Planetary mass ratio, $M_b/M_A$ ($\times 1000$) & $0.22$ & $0.18 $ & $-0.11$ & $+0.13$\\
\multicolumn{5}{c}{\bf Planetary Orbit} \\
Orbital Period, $P_b$ (day) & $ 105.595$ & $ 105.599 $ & $-   0.038$ & $+   0.053$\\
Eccentricity Parameter, $\sqrt{e_b} \cos(\omega_b)$ & $ 0.062$ & $ 0.046 $ & $- 0.064$ & $+ 0.049$\\
Eccentricity Parameter, $\sqrt{e_b} \sin(\omega_b)$ & $ 0.040$ & $ 0.004 $ & $- 0.100$ & $+ 0.106$\\
Time of Barycentric Transit, $t_b$ (days since $t_0$) & $ -37.888$ & $ -37.896 $ & $-   0.022$ & $+   0.044$\\
Orbital Inclination, $i_b$ (deg) & $89.446$ & $89.442 $ & $- 0.026$ & $+ 0.030$\\
Relative Nodal Longitude, $\Delta \Omega$ (deg) & $-0.012$ & $-0.005 $ & $- 0.052$ & $+ 0.050$\\
\multicolumn{5}{c}{\bf Stellar Orbit} \\
Orbital Period, $P_1$ (day) & $  18.79537$ & $  18.79535 $ & $-0.000051$ & $+0.000062$\\
Eccentricity Parameter, $\sqrt{e_1} \cos(\omega_1)$ & $ -0.0074$ & $ -0.0074 $ & $-  0.0002$ & $+  0.0002$\\
Eccentricity Parameter, $\sqrt{e_1} \sin(\omega_1)$ & $ -0.32113$ & $ -0.32266 $ & $-  0.00188$ & $+  0.00185$\\
Time of Primary Eclipse, $t_1$ (days since $t_0$) & $-17.127434$ & $-17.127462 $ & $-  0.000078$ & $+  0.000072$\\
Orbital Inclination, $i_1$ (deg) & $89.265$ & $89.256 $ & $- 0.025$ & $+ 0.026$\\
\multicolumn{5}{c}{\bf Radius/Light Parameters} \\
Linear Limb Darkening Parameter, $u_A$ & $ 0.453$ & $ 0.457 $ & $- 0.007$ & $+ 0.007$\\
Quadratic Limb Darkening Parameter, $v_A$ & $0.143$ & $0.135 $ & $-0.018$ & $+0.018$\\
Stellar Flux Ratio, $F_B/F_A$ ($\times 100$) & $0.09081$ & $0.09059 $ & $-0.00072$ & $+0.00071$\\
Radius of Star A, $R_A$ ($R_\odot$) & $1.757$ & $1.752 $ & $-0.034$ & $+0.031$\\
Radius Ratio, Star B, $R_B/R_A$ & $0.15503$ & $0.15513 $ & $-0.00021$ & $+0.00021$\\
Planetary Radius Ratio, $R_b/R_A$ & $0.02272$ & $0.02254 $ & $-0.00030$ & $+0.00030$\\
\multicolumn{5}{c}{{\bf Relative Contamination, $F_{\rm cont}/F_A$} ($\times 100$)} \\
Quarter 1 & & & 0 (fixed) &\\
Quarter 2 & $ 1.40$ & $ 1.37 $ & $- 0.14$ & $+ 0.14$\\
Quarter 3 & $ 1.55$ & $ 1.60 $ & $- 0.14$ & $+ 0.14$\\
Quarter 4 & $ 1.88$ & $ 1.91 $ & $- 0.14$ & $+ 0.14$\\
Quarter 5 & $ 0.58$ & $ 0.62 $ & $- 0.14$ & $+ 0.14$\\
Quarter 6 & $ 1.21$ & $ 1.17 $ & $- 0.14$ & $+ 0.14$\\
Quarter 7 & $ 1.48$ & $ 1.52 $ & $- 0.15$ & $+ 0.15$\\
Quarter 8 & $ 1.73$ & $ 1.77 $ & $- 0.14$ & $+ 0.14$\\
Quarter 9 & $ 0.78$ & $ 0.83 $ & $- 0.14$ & $+ 0.14$\\
Quarter 10 & $ 1.26$ & $ 1.25 $ & $- 0.14$ & $+ 0.14$\\
Quarter 11 & $ 1.80$ & $ 1.81 $ & $- 0.14$ & $+ 0.14$\\
\multicolumn{5}{c}{\bf Noise Parameter}  \\
Long Cadence Relative Width, $\sigma_{\rm LC}$ ($\times 10^5$) & $17.25$ & $17.19 $ & $- 0.16$ & $+ 0.16$\\
\multicolumn{5}{c}{\bf Radial Velocity Parameters} \\
RV Offset, $\gamma$ (km s$^{-1}$) & $18.008$ & $17.996 $ & $- 0.030$ & $+ 0.031$\\
Zero-level Diff., McDonald/HET, $\Delta \gamma$ (km s$^{-1}$) & $-0.083$ & $-0.058 $ & $- 0.035$ & $+ 0.034$\\
\enddata
\tablecomments{The reference epoch is $t_0=2,454,970$ (BJD).}
\end{deluxetable}

\clearpage

\begin{deluxetable}{lcccc}
\tablecaption{Derived
Parameters from Photometric Dynamical Model\label{derived}}
\tablewidth{0pt}
\tablehead{
\colhead{Parameter} &          
\colhead{Best fit}  &
\colhead{50\%} &
\colhead{15.8\%} &
\colhead{84.2\%} 
}
\startdata
Mass of Star A, $M_A$ ($M_\odot$) & $0.949$ & $0.941 $ & $-0.059$ & $+0.055$\\
Mass of Star B, $M_B$ ($M_\odot$) & $0.249$ & $0.248 $ & $-0.010$ & $+0.009$\\
Radius of Star A, $R_A$ ($R_\odot$) & $1.757$ & $1.752 $ & $-0.034$ & $+0.031$\\
Radius of Star B, $R_B$ ($R_\odot$) & $0.2724$ & $0.2717 $ & $-0.0053$ & $+0.0049$\\
Radius of Planet b, $R_b$ ($R_\oplus$) & $4.35$ & $4.30 $ & $-0.11$ & $+0.11$\\
Mass of Planet b, $M_b$ ($M_\oplus$) & \multicolumn{4}{l}{$ < 122$ (95\% conf.)} \\
Density of Star A, $\rho_A$ (g cm$^{-3}$) & $0.1749$ & $0.1750 $ & $-0.0014$ & $+0.0014$\\
Density of Star B, $\rho_B$ (g cm$^{-3}$) & $ 12.32$ & $ 12.35 $ & $-  0.25$ & $+  0.26$\\
Gravity of Star A, $\log g_A$ (cgs) & $  3.926$ & $  3.925 $ & $-  0.011$ & $+  0.010$\\
Gravity of Star B, $\log g_B$ (cgs) & $ 4.9640$ & $ 4.9635 $ & $- 0.0026$ & $+ 0.0026$\\
Fractional radius of Star A, $R_A/a$ & $0.05562$ & $0.05560 $ & $-0.00012$ & $+0.00011$\\
Fractional radius of Star B, $R_B/a$ & $0.008623$ & $0.008624 $ & $-0.000026$ & $+0.000026$\\
Semimajor Axis of Stellar Orbit, $a_1$ (AU) & $0.1469$ & $0.1466 $ & $-0.0029$ & $+0.0026$\\
Semimajor Axis of Planet b, $a_b$ (AU) & $0.4644$ & $0.4632 $ & $-0.0092$ & $+0.0082$\\
Eccentricity of Stellar Orbit, $e_1$ & $0.1032$ & $0.1042 $ & $-0.0012$ & $+0.0012$\\
Argument of Periapse Stellar Orbit, $\omega_1$ (deg) & $ 268.680$ & $ 268.695 $ & $-   0.039$ & $+   0.037$\\
Eccentricity of Planet b Orbit, $e_2$ & \multicolumn{4}{l}{$ < 0.032$ (95\% conf.)} \\
Mutual Orbital Inclination\tablenotemark{a}, $I$ (deg) & $0.182$ & $0.191 $ & $-0.032$ & $+0.037$\\
\enddata
\tablenotetext{a}{The mutual inclination is the angle between
the orbital planes of the binary and the planet, and is
defined as
$\cos I  =  \sin i_1 \sin i_b \cos \Delta \Omega 
+ \cos i_1 \cos i_b$}
\end{deluxetable}

\begin{deluxetable}{cc}
\tablecaption{Parameters from ELC Model\label{ELCparm}}
\tablewidth{0pt}
\tablehead{
\colhead{Parameter} &          
\colhead{Best fit}
}
\startdata
$K_A$ (km s$^{-1}$) & $17.794 \pm 0.031$ \\
$e$                 & $0.1030 \pm 0.0012$ \\
$\omega$ (deg)      & $268.68\pm 0.04$ \\
$R_A/a$             & $0.05493\pm       0.00023$ \\
$R_B/a$             & $0.00870\pm       0.00006$ \\
$T_{\rm eff, B}/T_{\rm eff, A}$           & $0.5896\pm        0.0026$  \\
$i$ (deg)           & $89.412   \pm    0.067$ \\
$x_A$               & $0.459\pm       0.027$ \\
$y_A$               & $0.130\pm       0.006$  \\
$P$ (days)          & $18.7952667\pm 0.0000020$ \\
$T_{\rm conj}$ (BJD)     & $2,454,971.66790\pm 0.00005$ \\
\enddata
\tablecomments{
Note: Subscript ``A'' denotes the primary star, subscript ``B'' the
secondary star.}
\end{deluxetable}

\end{document}